\pdfoutput=1
\documentclass[10pt,conference]{IEEEtran}

\usepackage{xparse}
\usepackage{xspace}
\usepackage{listings}
\usepackage{xcolor}
\usepackage{colortbl}
\usepackage{tikz}
\usepackage{arydshln}
\usepackage{booktabs}
\usepackage{balance}
\usepackage{dblfloatfix}
\usepackage{enumitem}
\usepackage{scalerel}
\usepackage{graphicx}
\usepackage{adjustbox}
\usepackage{quoting}
\usepackage[most]{tcolorbox}

\usepackage{amsmath}
\usepackage{amssymb}
\usepackage{amsfonts}

\usepackage{cite}

\usepackage[hidelinks]{hyperref}
\usepackage[capitalize,noabbrev]{cleveref}

\quotingsetup{
  leftmargin=0.8em,
  rightmargin=0.8em,
  vskip=0pt,
  font=itshape
}

\newtcolorbox{quotebox}{
  colback=gray!5,
  colframe=gray!50,
  boxrule=0.3pt,
  arc=1pt,
  left=6pt,
  right=6pt,
  top=4pt,
  bottom=4pt,
  before skip=6pt,
  after skip=6pt,
  fontupper=\small\itshape,
}

\providecommand{\citet}{}
\providecommand{\citep}{}

\newcommand{\toolname}{\textsc{Any}PoC}
\newcommand{\tool}{\toolname\xspace}
\newcommand{\toolclaude}{\toolname$_{\scriptscriptstyle\textit{Claude}}$\xspace}
\newcommand{\toolcodex}{\toolname$_{\scriptscriptstyle\textit{Codex}}$\xspace}

\newcommand{\eg}{\emph{e.g.,}\xspace}
\newcommand{\llm}{LLM\xspace}
\newcommand{\llms}{LLMs\xspace}
\newcommand{\llmbased}{LLM-based\xspace}
\newcommand{\llmagents}{LLM agents\xspace}
\newcommand{\poc}{PoC\xspace}
\newcommand{\pocs}{PoCs\xspace}
\newcommand{\sonnet}[1]{Sonnet #1\xspace}
\newcommand{\opus}[1]{Opus #1\xspace}
\newcommand{\gptcodex}[1]{GPT-#1-Codex\xspace}
\newcommand{\sut}{SUT\xspace}
\newcommand{\suts}{SUTs\xspace}
\newcommand{\cc}{Claude Code\xspace}
\newcommand{\codex}{Codex\xspace}
\newcommand{\simplereporter}{\textit{LLM-Reporter}\xspace}
\newcommand{\analyzer}{\textit{analyzer}\xspace}
\newcommand{\generator}{\textit{generator}\xspace}
\newcommand{\checker}{\textit{checker}\xspace}
\newcommand{\knowledgeExtractor}{\textit{knowledge extractor}\xspace}
\newcommand{\knowledgeFilter}{\textit{knowledge filter}\xspace}
\newcommand{\phase}[1]{Phase #1\xspace}
\newcommand{\markdown}{Markdown\xspace}

\renewcommand{\sectionautorefname}{\S\kern-\fontdimen2\font}
\renewcommand{\subsectionautorefname}{\S\kern-\fontdimen2\font}
\renewcommand{\subsubsectionautorefname}{\S\kern-\fontdimen2\font}

\newcommand{\parabf}[1]{\noindent\textbf{#1}}
\newcommand{\q}[1]{``#1''}
\let\oldcite\cite
\let\oldcitet\citet
\let\oldcitep\citep
\renewcommand{\cite}{\unskip~\oldcite}
\renewcommand{\citet}{\unskip~\oldcitet}
\renewcommand{\citep}{\unskip~\oldcitep}
\newcommand{\code}[1]{{\small \texttt{#1}}}
\renewcommand{\citet}{\cite}
\renewcommand{\citep}{\cite}

\title{\tool: Universal Proof-of-Concept Test Generation for Scalable LLM-Based Bug Detection}

\author{
  \IEEEauthorblockN{Zijie Zhao, Chenyuan Yang, Weidong Wang, Yihan Yang, Ziqi Zhang, Lingming Zhang}
  \IEEEauthorblockA{University of Illinois Urbana-Champaign, Champaign, USA\\
  \{zijie4, cy54, weidongw, yihany5, ziqi24, lingming\}@illinois.edu}
}

\newcommand{\numSUT}{12\xspace}
\newcommand{\campaignTotalRuns}{1214\xspace}
\newcommand{\campaignAllPoCsApproximate}{2700\xspace}

\newcommand{\datasetLLMPos}{48\xspace}
\newcommand{\datasetLLMNeg}{48\xspace}
\newcommand{\datasetHumanPos}{48\xspace}
\newcommand{\datasetHumanNeg}{48\xspace}
\newcommand{\datasetTotal}{192\xspace}
\newcommand{\datasetTotalTrue}{96\xspace}

\newcommand{\evalPosImprovePercentageClaude}{44\%\xspace}
\newcommand{\evalPosImprovePercentageCodex}{30\%\xspace}
\newcommand{\evalPosImprovePercentageBoth}{37\%\xspace}
\newcommand{\evalNegativeRejectImprovementBoth}{9.7$\times$\xspace}
\newcommand{\evalNegativeNotRejectedRateClaude}{96\%\xspace}
\newcommand{\evalNegativeNotRejectedRateCodex}{85\%\xspace}
\newcommand{\evalNegativeRejectRateOurClaude}{85\%\xspace}
\newcommand{\evalNegativeRejectRateOurCodex}{96\%\xspace}

\newcommand{\numTotalBugs}{121\xspace}
\newcommand{\numBugsFixed}{92\xspace}
\newcommand{\numBugsConfixed}{108\xspace}
\newcommand{\numPoCAsTest}{46\xspace}
\newcommand{\numBugsOverOneMillion}{80\xspace}

\newcommand{\kbCompNokbValid}{56\xspace}
\newcommand{\kbCompKbValid}{65\xspace}
\newcommand{\kbCompValidPct}{16\%\xspace}
\newcommand{\kbCompNokbUnique}{3\xspace}
\newcommand{\kbCompKbUnique}{12\xspace}
\newcommand{\kbCompNokbCost}{\$5.02\xspace}
\newcommand{\kbCompKbCost}{\$4.67\xspace}
\newcommand{\kbCompCostPct}{7\%\xspace}

\newcommand{\evalAgentFalseTotalAvg}{87\xspace}
\newcommand{\evalAgentAnalyzerFalseAvg}{62.5\xspace}
\newcommand{\evalAgentAnalyzerFalsePctAvg}{71.8\%\xspace}
\newcommand{\evalAgentGeneratorFalseAvg}{18\xspace}
\newcommand{\evalAgentGeneratorFalsePctAvg}{20.7\%\xspace}
\newcommand{\evalAgentCheckerFalseAvg}{6.5\xspace}
\newcommand{\evalAgentCheckerFalsePctAvg}{7.5\%\xspace}

\newcommand{\priceMatchBaselineGenerated}{366\xspace}
\newcommand{\priceMatchOursGenerated}{82\xspace}
\newcommand{\priceMatchTriageRatio}{4.5$\times$\xspace}

\newcommand{\kbEffNumBugs}{53\xspace}
\newcommand{\kbEffToolCallsPct}{20\%\xspace}
\newcommand{\kbEffBashPct}{33\%\xspace}
\newcommand{\kbEffNokbRereadsAvg}{3.6\xspace}
\newcommand{\kbEffKbRereadsAvg}{1.8\xspace}

\newcommand{\kbTotalItems}{340\xspace}
\newcommand{\kbTotalUsages}{808\xspace}
\newcommand{\kbAvgItemsPerGen}{2.4\xspace}
\newcommand{\kbAvgRating}{6.0\xspace}
\newcommand{\kbMedianRating}{7.0\xspace}
\newcommand{\kbTotalVersionUpdates}{135\xspace}
\newcommand{\kbAvgRatingImprovementPerUpdate}{0.30\xspace}

\begin{document}

\maketitle

\begin{abstract}
While recent LLM-based agents can identify many candidate bugs in source code, their reports remain static hypotheses that require manual validation, limiting the practicality of automated bug detection.
We frame this challenge as a test generation task: given a candidate report, synthesizing an executable proof-of-concept (\poc) test, or simply a PoC --- such as a script, command sequence, or crafted input --- to trigger the suspected defect.
Automated \poc generation can act as a scalable validation oracle, enabling end-to-end autonomous bug detection by providing concrete execution evidence.
However, naive \llm agents are unreliable validators: they are biased toward \q{success} and may reward-hack by producing plausible but non-functional \pocs or even hallucinated traces.

To address this, we present \tool, a general multi-agent framework that
(1) analyzes and fact-checks a candidate bug report,
(2) iteratively synthesizes and executes a \poc while collecting execution traces, and
(3) independently re-executes and scrutinizes the \poc to mitigate hallucination and reward hacking.
In addition, \tool also continuously extracts and evolves a \poc knowledge base to handle heterogeneous tasks.
\tool operates on candidate bug reports regardless of their source and can be paired with different bug reporters.
To demonstrate practicality and generality, we apply \tool, together with a simple agentic bug reporter, on \numSUT large-scale, critical software systems, including Firefox, Chromium, LLVM, OpenSSL, SQLite, FFmpeg, and Redis.
Compared to the state-of-the-art coding agents, \eg \cc and \codex,
\tool produces \evalPosImprovePercentageBoth more valid \pocs for true-positive bug reports and rejects \evalNegativeRejectImprovementBoth more false-positive bug reports.
\tool also enables the discovery of \numTotalBugs new bugs from over two thousand noisy bug reports, with \numBugsConfixed confirmed by developers and \numBugsFixed fixed.
\numPoCAsTest \pocs have also been adopted as official regression tests.
\end{abstract}

\section{Introduction}
\label{sec:intro}

Automated bug detection based on Large Language Models (\llms) has been extensively studied in recent years\cite{anthropic2025securityreview,openai2026codexsecurity,GoogleBigSleep,wu2025bughundredsbehindllms,cursor2026bugbot,zhou2024large,du2024vul,knighter}.
Different from traditional static analysis\cite{clang-static-analysis,codeql,semgrep} and dynamic testing\cite{oss-fuzz,syzkaller,fraser2011evosuite,pacheco2007randoop}, recent \llmagents can autonomously explore the codebase and detect bugs in large systems.
Those \llmbased bug detection tools from both industry
\cite{GoogleBigSleep,openai2025aardvark,cursor2026bugbot,anthropic2025securityreview,github2025copilotreview}
and academia
\cite{guo2025repoaudit,wu2025bughundredsbehindllms}
have shown great promise by finding hundreds of bugs and security vulnerabilities in real-world software systems.

\begin{figure}[!t]
\centering
\includegraphics[width=\columnwidth]{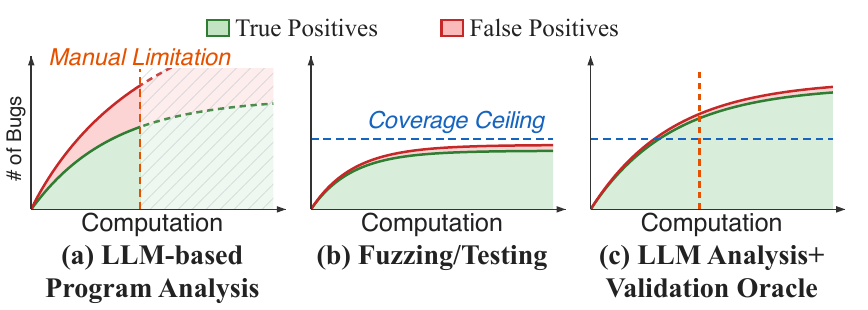}
\caption{Computation scalability of bug finding.
}
\label{fig:computation}
\vspace{-0.8em}
\end{figure}

Existing \llmbased bug detection techniques often suffer from high false positive rates\cite{guo2025repoaudit,wu2025bughundredsbehindllms,kaplan2026bugbot}.
\llmagents may fail to gather the full context, misinterpret code semantics, or hallucinate, leading to invalid bug reports.
While there are tools to mitigate this issue by using another model to validate the reports\cite{li2024enhancing}, the final outputs typically remain textual bug reports rather than concrete evidence.
Thus, developers often need to manually verify whether the reports are real, which is difficult and time-consuming\cite{du2026reducingfalsepositivesstatic,anthropic2026claudefirefox}.
This human-validation bottleneck fundamentally limits the scalability of \llmbased bug detection.
\autoref{fig:computation}(a) shows true bugs (green area) and false bugs (red area) growing with computation; the human-validation limit (orange line) prevents \llmbased detection from scaling to more bugs (shaded area).

In contrast, dynamic testing and fuzzing techniques\cite{oss-fuzz,syzkaller,pacheco2007randoop} rely on concrete executions and oracles (such as crashes and sanitizers\cite{serebryany2012addresssanitizer,song2019sok,stepanov2015memorysanitizer}) to reliably flag bugs without manual verification. For example, scalable fuzzing campaigns can discover tens of thousands of bugs~\cite{oss-fuzz}.
As shown in \autoref{fig:computation}(b), strong oracles allow fuzzing to scale to a large amount of computation with low false positive rates (the small red area). In recent years, \llmbased fuzzing techniques \cite{kernelgpt,xia2024fuzz4all,deng2023large,meng2024large,ou2024mutators} have also been widely studied, and can substantially complement traditional fuzzers.
However, we are limited by the fuzzing ceiling as shown in \autoref{fig:computation}(b).
This ceiling arises from several factors: fuzzers typically rely on a restricted set of runtime oracles, can only exercise modules that are available and runnable, and must explore an often enormous input and path space. In practice, especially for large systems with complex configurations, environment, or hardware dependencies, the fuzzed binary covers only a small fraction of the full codebase. As a result, even with more computation or LLM assistance, dynamic testing reaches diminishing returns as new bug-triggering paths become harder to discover.

Our insight is to combine the advantages of both lines.
\llmbased analysis for bug detection does not have a plateau because it can continuously cover more paths and edge cases through abstract reasoning.
Fuzzing/testing does not rely on manual verification and can scale to larger computation because of strong automated oracles.
We want to use strong and automatic oracles to validate the bugs from \llmbased detection systems so that we can effectively scale computation to more bugs.
As shown in \autoref{fig:computation}(c), the number of bugs is not limited by the fuzzing ceiling nor human effort.

To this end, we cast bug report validation as a test-generation problem: synthesizing an executable artifact that reliably triggers the suspected defect in a candidate report.
We call this artifact a proof-of-concept (\poc) test, or simply a PoC.
A \poc can take different forms, such as a standalone script, command sequence, or crafted input, depending on the target system and bug type.
By providing concrete execution-based evidence rather than a textual hypothesis, \pocs substantially reduce the manual effort to verify reported bugs and thus serve as a practical validation oracle for \llmbased bug detection.
Recent work has started to explore test or \poc generation from bug reports, including agentic techniques \cite{ahmed2025otter,marques2025explodejs,simsek2025pocgengeneratingproofofconceptexploits,nitin2025faultline}. However,
reliably turning \poc generation into a general-purpose validation oracle remains challenging (\autoref{sec:background:problem-setting}):
\begin{itemize}[leftmargin=*,topsep=0pt]
    \item \textbf{C1: \llm reward hacking.}
        \llms are rewarded to produce outputs that appear to be successful, even if the outputs are incorrect. This usually leads to hallucinated or non-functional \pocs.
        As a validation oracle, a bad \poc is worse than no output, so the system must enforce either high-quality \pocs or faithful rejection.
    \item \textbf{C2: Heterogeneity.}
        Different systems require vastly different types of \pocs.
        A browser bug may need a multi-process script driving DevTools, while a compiler bug may need a crafted IR input.
        Constructing a valid \poc in the correct format that genuinely demonstrates the bug is itself non-trivial.
    \item \textbf{C3: Scalability.}
        Generating valid \pocs requires extensive exploration of the codebase, build system, and tooling.
        Repeating this for every bug in a large codebase is costly without a way to accumulate and reuse knowledge.
\end{itemize}

To tackle these challenges, we present \tool, a novel multi-agent framework for automated \poc generation with a self-evolving knowledge base (KB), detailed in \autoref{sec:approach}.
We carefully decompose the \poc generation task into several dedicated subagents: analyzer, generator, validator, and knowledge extractor. Each subagent specializes in different aspects of the \poc generation process.
Using a candidate bug report as input, \tool first validates factual correctness of the bug report,
then generates an executable \poc together with execution traces,
which would be rigorously re-executed and examined by a validator subagent to avoid hallucination and invalid assumptions.
Moreover, we design agents to automatically extract, query, and evolve a knowledge base about the target codebase during the \poc generation process for scalability.
This allows \tool to continuously learn and adapt to the unique characteristics of different codebases without relying on hard-coded rules or patterns.
\tool is designed to be a general-purpose validation layer that works with any bug reporter, including \llmbased tools, traditional checkers, or even reports from humans.
\tool is bug-type-, project-, and oracle-agnostic, allowing it to be applied to a wide range of software systems and bugs without special customization.

We evaluate \tool on \numSUT diverse real-world software systems, including Chromium, Firefox, OpenSSL, SQLite, and FFmpeg.
The results in \autoref{sec:eval} show that, compared with state-of-the-art baseline agents, \eg \cc~\cite{claudecode} and \codex~\cite{codex}, \tool produces \evalPosImprovePercentageBoth more valid \pocs for true-positive bug reports while rejecting \evalNegativeRejectImprovementBoth more false-positive reports.
To further demonstrate its practical value, we pair \tool with a simple \llm-based bug reporter for new bug discovery on these systems.
To date, \tool has already detected \numTotalBugs new bugs, with \numBugsConfixed confirmed by developers, \numBugsFixed fixed, and \numPoCAsTest generated \pocs adopted as regression tests.
These results highlight the promising future of combining \llmbased bug detection and \poc generation as a powerful paradigm for automated software quality assurance.
In summary, our main contributions are:
\begin{itemize}[leftmargin=*,topsep=2pt]
    \item \textbf{Universal \poc generation framework for bug report validation.}
        We design a universal and scalable multi-agent \poc generation system, \tool, for validating textual bug reports,
        enabling automated bug detection to scale to a large amount of computation without human intervention.
    \item \textbf{Comprehensive Evaluation:}
        We conduct extensive experiments on \numSUT real-world software systems (\eg Chromium, Firefox, LLVM, OpenSSL, FFmpeg, and Redis) to show the effectiveness of \tool over state-of-the-art coding agent systems, \eg \cc and \codex.
    \item \textbf{Real-world impact:}
        On those complex and fundamental real-world systems, \tool enabled the discovery of \textbf{\numTotalBugs} new bugs, with \textbf{\numBugsConfixed} confirmed and \textbf{\numBugsFixed} fixed. Notably, \numBugsOverOneMillion of the bugs were found in projects with over one million lines of code. Moreover, \textbf{\numPoCAsTest} generated \pocs have been adopted as official regression tests.
\end{itemize}

\section{Background and Related Work}
\label{sec:background}

\begin{table*}[t]
\centering
\caption{Comparison of \tool with prior test/\poc generation approaches.}
\label{tab:tool_comparison}
\small\setlength{\tabcolsep}{4pt}
\begin{tabular}{l@{\hspace{1.5em}}ccccccc}
\toprule
& \textbf{Language} & \textbf{External Dependency} & \textbf{Bug Type} & \textbf{PoC Type} & \textbf{Scaffold} & \textbf{FP Filtering} & \textbf{New Bugs} \\
\midrule
Libro \cite{kang2023libro}                                    & Java         & javalang \cite{javalang2020}        & Any       & Unit test & N/A   & {\color{red}$\times$} & {\color{red}$\times$}   \\
Otter \cite{ahmed2025otter}                                    & Python       & Flake8 \cite{flake8}                & Any       & Unit Test      & N/A & {\color{red}$\times$} & {\color{red}$\times$}   \\
Explode.js \cite{marques2025explodejs}                        & JavaScript   & Graph.js \cite{graphjs}, Z3 \cite{z3} & 4 classes & JavaScript Program & N/A   & {\color{red}$\times$} & 44   \\
PoCGen \cite{simsek2025pocgengeneratingproofofconceptexploits} & JavaScript   & CodeQL \cite{codeql}                & 5 classes & JavaScript Program   & Fixed & {\color{red}$\times$} & {\color{red}$\times$}   \\
SmartPoC \cite{chen2025smartpoc} &  Solidity  &   Foundry \cite{foundry}       & Any & Smart Contract   & Fixed & {\color{red}$\times$} & {\color{red}$\times$}   \\
FaultLine \cite{nitin2025faultline}                        & Python, Java & None              & 4 classes & Any       & Fixed & {\color{red}$\times$} & {\color{red}$\times$}   \\ \hline
\textbf{\tool}                                      & \textbf{Any} & \textbf{None}     & \textbf{Any} & \textbf{Any} & \textbf{Any} & \textbf{{\color{green!60!black}$\checkmark$}} & \textbf{\numTotalBugs} \\
\bottomrule
\end{tabular}
\vspace{-0.8em}
\end{table*}

In this section, we first introduce existing bug detection techniques based on program analysis and their shared limitations, and then discuss existing work on \poc generation.

\subsection{Program Analysis for Bug Detection}
\label{sec:background:static-bug-detection}

Traditional static analysis is widely used for scalable bug detection\cite{clang-static-analysis,codeql,semgrep}, but often faces high false positive rates, restricted bug coverage, and substantial manual effort required to construct specifications, rules, or checkers. As a result, existing analyzers often struggle to detect diverse and complex bugs in large real-world codebases\cite{knighter}.

Large language models (\llms) offer a promising way to address these limitations. By leveraging rich semantic knowledge and contextual reasoning, \llms generalize beyond manually encoded rules and support more flexible code analysis, making \llmbased bug detection an increasingly promising direction for continuous and scalable code analysis.
In academia, researchers have explored a range of \llm-powered approaches, including inferring taint specifications for external APIs (IRIS\cite{li2024iris}, Artemis\cite{ji2025artemis}), synthesizing static analysis checkers (KNighter\cite{knighter}, QLCoder\cite{wang2025qlcoder}), and directly scanning source code for vulnerabilities (Vul-RAG\cite{du2024vul}, RepoAudit\cite{guo2025repoaudit}).
Many companies have also developed various \llmbased or agentic tools for code review, bug detection, and security auditing.
Google's Big Sleep project has found vulnerabilities in real-world codebases\cite{GoogleBigSleep}.
GitHub's Copilot Code Review \cite{github2025copilotreview} and Cursor's Bugbot \cite{cursor2026bugbot} are two widely-used \llmbased tools for automated bug detection. More recently, OpenAI provides Codex Security \cite{openai2026codexsecurity,openai2025aardvark} to find complex vulnerabilities, while Claude Code also supports security review \cite{anthropic2025securityreview} and code review \cite{anthropic2026codereview}.

Despite this promise, current \llmbased systems often generate many candidate reports that remain unverified hypotheses rather than confirmed bugs. Because \llms are probabilistic and may misinterpret complex code semantics, execution conditions, or bug consequences, their outputs can contain substantial false positives\cite{wu2025bughundredsbehindllms,knighter}. As a result, developers must manually inspect each report to determine whether it is real, reproducible, and security-relevant. This process is costly, time-consuming, and requires significant domain expertise, making it a major bottleneck at scale.
Therefore, a key next step is automated validation that can efficiently and reliably confirm reported bugs.

\subsection{Test and \poc Generation}
\label{sec:background:poc-generation}

Test and \poc generation can automatically validate bug reports.
Libro\cite{kang2023libro} used \llms to generate bug-reproducing unit tests from bug reports for the Defects4J benchmark \cite{just2014defects4j}, but it is limited to the JUnit format.
\citet{plein2024automatic} studied the feasibility of \llm-based bug-report-to-test generation, while \citet{qureshi2025test} further evaluated the robustness of this setting via a cognitive-layered evaluation.
Recently, Otter \cite{ahmed2025otter} generates fail-to-pass tests from issue descriptions, with the explicit goal of validating candidate patches.
More broadly, recent general software engineering agents such as SWE-agent~\cite{yang2024sweagent} and Agentless~\cite{xia2024agentless} also incorporate reproduction test generation as part of issue resolution, using such tests to help confirm bugs and assess proposed fixes.
While these techniques are useful for regression testing and patch validation, they generally rely on internal APIs, test harnesses, or developer-side infrastructure.
Thus, they are not designed to produce standalone evidence that a bug can be triggered across diverse realistic settings.

Similarly, given a candidate bug report, \poc generation aims to automatically synthesize an executable artifact that directly triggers the reported bug or vulnerability, providing concrete evidence that lets developers quickly validate it.
However, existing \poc generation works\cite{marques2025explodejs,simsek2025pocgengeneratingproofofconceptexploits,nitin2025faultline} have two limitations.
First, they lack scalability across a wide range of programming languages, bug report formats, or \poc formats.
Second, they are evaluated on datasets with only true bugs and do not explicitly handle false reports.
\autoref{tab:tool_comparison} shows the comparison of \tool with prior approaches.
Explode.js\cite{marques2025explodejs} is limited to JavaScript and a fixed set of common bug classes for Node.js.
PoCGen\cite{simsek2025pocgengeneratingproofofconceptexploits} handles five vulnerability types in npm packages and relies heavily on the setup of CodeQL.
More recently, SmartPoC\cite{chen2025smartpoc} uses a fixed framework to generate \poc from smart contract bug reports, while
FaultLine\cite{nitin2025faultline} supports four bug classes where dataflow analysis is necessary.
In summary, existing work is insufficient for general-purpose bug-report validation due to scalability issues and limitations in handling false reports.

\begin{figure*}[!t]
\centering
\includegraphics[
    width=0.85\linewidth,
    keepaspectratio,
    clip
]{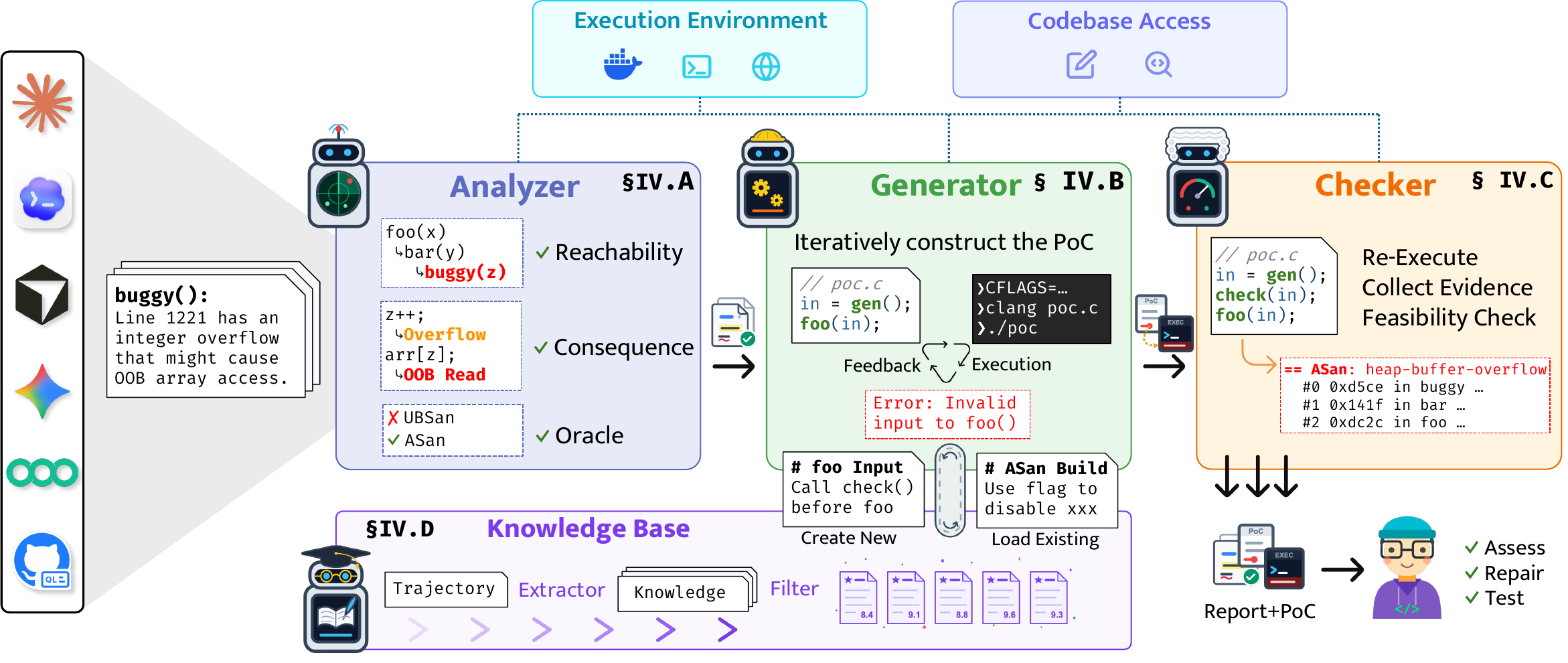}
\caption{Overview of the \tool framework.}
\label{fig:overview}
\vspace{-0.8em}
\end{figure*}

\section{Problem Setting}
\label{sec:background:problem-setting}

After identifying the limitations of existing work,
we now define the problem to clarify the scope and requirements of our work.
Let $\mathcal{S}$ denote a system, which can be a software application, library, or any executable codebase.
Let $\mathcal{R}$ denote a bug reporter that can produce many candidate bug reports $\mathcal{R}(\mathcal{S})$ for $\mathcal{S}$.
Each candidate report $ r\in \mathcal{R}(\mathcal{S})$ describes a potential defect in $\mathcal{S}$ and may include information about symptoms, conditions, locations, and consequences.
A validator $\mathcal{V}$ takes as input a candidate bug report $r$ and the system $\mathcal{S}$, and outputs either a valid \poc $p_r$ or $\bot$,
where $\bot$ indicates that $r$ is invalid or that a valid $p_r$ cannot be generated.

{%
\setlength{\abovedisplayskip}{-4pt}%
\setlength{\belowdisplayskip}{4pt}%
\setlength{\abovedisplayshortskip}{2pt}%
\setlength{\belowdisplayshortskip}{2pt}%
\[
\mathcal{V}(\mathcal{S}, r) =
\begin{cases}
p_r, & \text{if $r$ is valid and \poc is possible},\\
\bot, & \text{otherwise}.
\end{cases}
\]
}Here, $p_r$ is an executable artifact that triggers the bug report $r$ in $\mathcal{S}$ and demonstrates its existence.
The goal of our work is to design a general and automated validator $\mathcal{V}$ that can efficiently and accurately generate $p_r$ for a wide range of $\mathcal{S}$ and $r \in \mathcal{R}(\mathcal{S})$, without relying on manual effort.
$\mathcal{V}$ should also faithfully output $\bot$ when the candidate bug report is invalid or when a valid $p_r$ cannot be generated.
We assume that the existence of a valid $p_r$ implies the validity of the corresponding $r$, but not vice versa.

The following are explicitly out of scope.
(1) We do not aim to improve the bug reporter $\mathcal{R}$; \tool is designed to generalize across different reporters, and we assume that with more computational resources, $\mathcal{R}$ produces more candidate reports, including more valid ones.
(2) We do not aim to prove the non-existence of a bug, which is generally very difficult in practice for large systems\cite{dijkstra1970structured,o2019incorrectness}.

We identify the goal of the validator $\mathcal{V}$ and its generated proofs-of-concept $p_r$ as follows:

\parabf{G1: Generality.} $\mathcal{V}$ should be able to handle arbitrary systems $\mathcal{S}$ and candidate bug reports $r \in \mathcal{R}(\mathcal{S})$,
without being tailored to specific bug types or software domains.
This requires the generated $p_r$ to be heterogeneous and not rely on a single format or oracle.
This is a key requirement that existing specialized \poc generation efforts do not address.

\parabf{G2: Automation.} $\mathcal{V}$ should operate without human intervention and with only minimal setup effort. In this way, $\mathcal{V}$ can scale to a large number of systems.

\parabf{G3: Actionability.} To reduce human effort, each $p_r$ must contain definitive evidence of the manifestation of the bug $r$.
A valid $p_r$ should exercise the system $\mathcal{S}$ through realistic usage scenarios and trigger the bug $r$.
This helps developers to quickly understand the bug and its severity.

\section{Approach}
\label{sec:approach}
\label{sec:approach:overview}

To address the challenges in \autoref{sec:intro} and satisfy the design requirements in \autoref{sec:background:problem-setting},
we design \tool to be a universal multi-agent framework that turns automated proof-of-concept (\poc) generation into a reliable oracle for validating textual bug reports.
\autoref{fig:overview} illustrates the overview of \tool.
We decompose the \poc generation task into three dedicated subagents --- \analyzer, \generator, and \checker.
Starting with bug reports from any given bug detection tool, the \analyzer subagent first statically examines a report to filter out clearly invalid ones and summarizes the findings.
Then, the \generator subagent iteratively produces a \poc to trigger the bug and passes it, with execution evidence, to the \checker subagent.
Finally, the \checker independently re-executes the \poc and verifies whether the evidence proves the bug's existence.
As shown in \autoref{fig:overview}, the \generator has access to a self-evolving knowledge base (KB) that accumulates reusable knowledge across \poc generation attempts.
A \knowledgeExtractor subagent distills useful knowledge from the \generator's trajectory and a \knowledgeFilter subagent validates such knowledge before it is committed to the KB.
Any future \generator can leverage prior experience, avoiding redundant exploration.
All subagents are equipped with a common set of tools, including an editor, search utilities, and bash execution.

\subsection{Bug Analysis}
\label{sec:approach:analysis}

When presented with a candidate bug report, \tool first conducts a thorough analysis before attempting \poc generation.
False bug reports might contain factual errors, such as incomplete code analysis, wrong assumptions about system behavior, or overstated consequences.
These errors could surface during \poc generation, but a separate analysis step has two advantages:
(1) it quickly filters out false reports before the more costly \poc generation step;
(2) the detailed analysis requires extensive exploration that would otherwise consume a large portion of the limited context window.
By separating the task, the generator can focus on \poc generation and benefit from the analysis summary.

\begin{figure}[t]
\centering
\includegraphics[width=\linewidth]{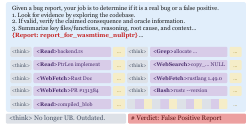}
\caption{Example trajectory of the bug analysis subagent.}
\label{fig:analyzer_prompt}
\vspace{-0.8em}
\end{figure}

\autoref{fig:analyzer_prompt} shows an example trajectory of the analyzer subagent.
Similar to all figures showing trajectories in the rest of this section, the grey box represents thinking or reasoning; the purple box represents tool invocation; and the yellow box represents environment feedback.
As shown in the figure, the analyzer is given the bug report and asked to understand the bug mechanism, including the root cause, consequence, and required oracle to demonstrate the bug manifestation.
The analyzer then determines the validity of the bug report based on all the collected context information and its reasoning about the bug mechanism.
The example in \autoref{fig:analyzer_prompt} is a false bug report for Wasmtime \cite{wasmtime2026} claiming a call to \code{ptr::copy\_nonoverlapping(src, dst, size)} with a \code{nullptr} and size being zero is undefined behavior (UB) in Rust.
The analyzer first examined the codebase to confirm the presence of such a call with the claimed parameters.
Then, while verifying the claimed UB consequence, the agent fetched two versions of Rust documentation using the web search tool.
The analyzer noticed a discrepancy between two versions and realized that the semantics of the suspected function have changed.
Finally, the analyzer confirmed the semantic change and correctly rejected this false report.
This example shows the analyzer can conduct thorough analysis and fact-checking, significantly reducing the burden on subsequent \poc generation both by filtering out false reports and by saving context-window space.

\subsection{PoC Generation}
\label{sec:approach:generation}

If the bug report passes the analysis step, \tool then spawns a generator subagent to produce a \poc for the candidate bug.
We design the generation to include two main phases.
During each phase, the generator can freely think and interact with the environment before finally deciding to stop.
This design ensures the generator can focus on one important aspect of the task at a time.

\begin{figure}[t]
\centering
\includegraphics[width=\linewidth]{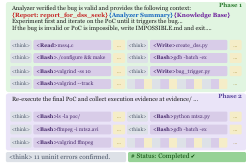}
\caption{Example trajectory of the PoC generator subagent.}
\label{fig:generator_prompt}
\vspace{-0.8em}
\end{figure}

\autoref{fig:generator_prompt} illustrates the workflow of a generator subagent.
In the prompt, we provide three pieces of contextual information:
(1) the summary of the bug mechanism from the analyzer subagent;
(2) the bug report itself;
(3) a snapshot of the knowledge base, which will be detailed later in \autoref{sec:approach:kb}.

In \phase{1}, the generator subagent is asked to experiment with \poc generation and iterate on the \poc until it successfully triggers the bug.
In the example, the subagent explores the codebase, writes a \poc program, and uses \texttt{valgrind}\cite{valgrind} and \texttt{gdb}\cite{gdb_manual} to perform multiple rounds of debugging.
The generator is instructed to extract a minimal \poc into a separate directory once it can trigger the bug.
Although a faithful agent should have already executed the \poc, we observe cases where the generator fails to do so correctly and finishes prematurely.
To mitigate reward hacking and hallucination, we enter \phase{2} and send another prompt asking the generator to execute the generated \poc again and collect execution traces into a dedicated evidence directory.
As shown in \autoref{fig:generator_prompt}, the generator carefully re-executes the final \poc and saves logs at the dedicated evidence directory.
After this phase, the generator finishes preparing all the necessary artifacts for the subsequent validation step.

During both \phase{1} and \phase{2}, we instruct the generator to give up on the task if it thinks the bug report is invalid, or \poc generation is impossible under the current environment.
If the generator gives up, it will output a short explanation and we will discard the bug report without further processing.

\subsection{PoC Validation}
\label{sec:approach:validation}

A generated \poc alone is not sufficient to confirm a bug: the generator may hallucinate execution results, produce a \poc that only appears to trigger the bug, or collect misleading evidence.
Even when the execution triggers the bug, the \poc may rely on unrealistic assumptions or an impractical setup (e.g., unrealistic flags), making it less useful for developers.
To guard against these failure modes, \tool employs a separate evidence checker subagent that independently re-executes and validates the \poc in a fresh environment.
Crucially, the checker operates without access to the generator's context window, ensuring that its judgment is not biased by the generator's reasoning.

\begin{figure}[t]
\centering
\includegraphics[width=\linewidth]{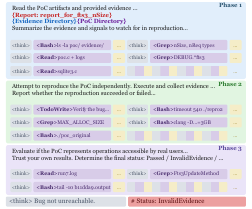}
\caption{Example trajectory of the evidence checker subagent.}
\label{fig:checker_prompt}
\vspace{-0.8em}
\end{figure}

\autoref{fig:checker_prompt} shows the prompt and an example workflow for the evidence checker subagent.
The validation process is composed of three phases.
In \phase{1}, the checker is given the bug report with the \poc artifacts and the execution evidence produced by the generator.
The checker is instructed to summarize what signals should be observed during the reproduction as an indication of the bug manifestation.
The signal can be any observable output such as a sanitizer crash or the difference between two execution traces.
In \autoref{fig:checker_prompt}, the bug report is about truncating a 64-bit number to a 32-bit number, thus potentially causing overflow.
The checker reads the \poc and execution logs and determines the signal for this bug would be an Undefined Behavior Sanitizer error.
Then, in \phase{2}, the checker is asked to copy the \poc artifacts into a fresh workspace, independently re-execute the \poc, and collect new execution traces.
The checker in the example executes the \poc in several different settings, but fails to observe any expected signal.
Finally, in \phase{3}, we ask the checker to check whether the new traces contain the expected signal and thus determine whether the \poc successfully triggers the bug.
We explicitly instruct the checker to trust its own execution results over any evidence claimed by the generator.
Here, since the checker could not trigger the bug in \phase{2}, it carefully checks the execution logs and finally concludes the \poc as invalid.

Beyond confirming the traces, we also ask the checker to evaluate whether the \poc represents a realistic scenario that real users or attackers could trigger through normal operations.
This filters out \pocs that rely on unrealistic assumptions or impractical setups.
Finally, if the checker determines that the \poc is valid, \tool will output the bug report, the \poc artifacts, and the execution evidence as the final output for human review.

\subsection{Self-Evolving Knowledge Base}
\label{sec:approach:kb}
To let the generator generate complex \pocs in large codebases like Firefox and Chromium,
we also build a self-evolving knowledge base (KB) that accumulates reusable knowledge across \poc generation attempts.
With the KB, the generator need not repeatedly explore the codebase to understand project-specific tools and caveats.

\parabf{Structure.}
The knowledge base is directory-based: each project has its own subdirectory, and each knowledge category its own subdirectory within it.
Each knowledge entry is stored as a \markdown file in the corresponding category subdirectory.
Each knowledge entry has four key properties:
(1) core content,
(2) keywords,
(3) usefulness rating,
(4) version number.
The core content and keywords are the content of a \markdown file, where we do not enforce a specific format to allow for flexibility.
The usefulness rating is a score from -10 to 10 that indicates how useful the entry is for \poc generation.
Each generator can provide a rating for any knowledge item. We keep all the historical ratings for each item so that future generators can better judge which knowledge item to use.
The version number is used to track the evolution of the entry.

From our empirical experience during bug finding, we design a few high-level knowledge categories that are most useful for \poc generation.
Namely, the categories include \textit{Command Line Tools}, \textit{Build System}, \textit{Internal Tools}, \textit{Test Frameworks}, \textit{Code}, and \textit{\poc Format}.
The \textit{Command Line Tools} category is shared across different projects, while the others are project-specific.
The \textit{Build System}, \textit{Internal Tools}, \textit{Test Frameworks} categories contain knowledge about project-specific tooling.
The \textit{Code} category is used to save understanding of code pieces that are heavily reused in the codebase.
The \textit{\poc Format} category directly keeps the possible types of \pocs.

\begin{figure}[t]
  \centering
  \includegraphics[width=\linewidth]{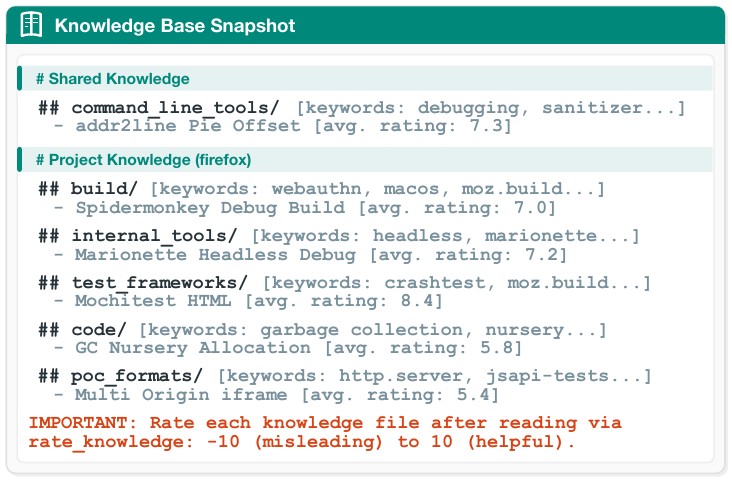}
  \caption{An example knowledge base snapshot.}
  \label{fig:kb_snapshot}
  \vspace{-0.8em}
\end{figure}

\parabf{Snapshot and Usage.}
As mentioned earlier in \autoref{sec:approach:generation}, the generator is given information about the current knowledge base.
We provide a succinct snapshot of the KB with top-rated knowledge items in each category, together with their average historical ratings and keywords.
As shown in \autoref{fig:kb_snapshot}, the snapshot lists one example knowledge item title for each category.
We also provide instructions for the generator to explore and search the KB content directories to find the content \markdown files.
To help future \poc generation, the generator is also explicitly instructed to provide a new rating for each knowledge item it accessed.

\parabf{Extraction and Evolution.}
By default, \tool starts working on a project with an empty initial knowledge base.
It builds up and evolves the knowledge base on the fly during \poc generation.
After each \poc generation attempt, whether it succeeds or fails, we task a knowledge extractor subagent to examine the generator's trajectory.
The extractor is given a short, index-based representation of the trajectory, with tools to inspect each step on demand.
If it notices reusable knowledge, we instruct it to either create new knowledge or update existing knowledge, so that an incorrect or incomplete item can later be fixed or improved.
Both new knowledge creation and updates are implemented as custom tools.

\parabf{Filtering.}
During implementation, we found that frontier models tend to report knowledge too specific to the particular bug, which is less reusable, and sometimes place entries in wrong categories, making them hard to find later.
To counter this, we introduce a simple filter subagent.
Whenever the extractor reports a knowledge item, the filter determines whether it is reusable and correctly categorized.
If it notices problems, it provides feedback, and the extractor can report again with updated content or drop this piece of knowledge.

\section{Evaluation}
\label{sec:eval}

To evaluate the effectiveness of \tool, we answer the following research questions:
\begin{itemize}[leftmargin=*,topsep=0pt]
    \item RQ1: Can \tool meet the generality requirements (\autoref{sec:background:problem-setting}) and enable scalable bug detection in practice?
    \item RQ2: How effective is \tool in being used as a validation oracle for \llm-based bug detection?
    \item RQ3: How does each component in \tool contribute to its effectiveness?
\end{itemize}

\subsection{Experimental Setup}
\label{sec:eval:bug_setup}

\parabf{Bug Reporter.}
As \poc generation approaches (both \tool and baselines) require a bug report as input, we implement a basic agentic bug reporter following the recent KNighter work~\cite{knighter} in our evaluation.
We refer to it as \simplereporter in this paper and use it across \tool and baselines for a fair comparison.
The reporter uses historical bug-fixing commits as bug patterns to find additional bugs.
It first selects commits within a time window, then iteratively constructs patterns from the commits.
For each pattern, the reporter searches for similar buggy code in the codebase and generates candidate bug reports.
Improving the reporter to generate more candidate reports is orthogonal to \tool and left as future work.
In our evaluation, \simplereporter already produces enough true bugs to study \tool's effectiveness.

\parabf{Systems Under Test (\suts).}
We evaluate \tool on real-world and large-scale systems.
We select \numSUT software across different domains that are widely used by billions of users across the world.
The systems are listed in \autoref{tab:projects}.
Nine of these systems have over 10K stars on GitHub and {six of the systems have over one million lines of code}.
The selected systems have also been widely studied by earlier software testing and analysis work\cite{hazimeh2020magma, klees2018evaluating, metzman2021fuzzbench, wang2025cybergym, will2012intoverflow}.

\begin{table}[t]
\centering
\caption{Systems under test in our bug finding campaign.}
\label{tab:projects}
\small\setlength{\tabcolsep}{4pt}
\adjustbox{max width=\linewidth}{
\begin{tabular}{llrrl}
\toprule
\textbf{Project} & \textbf{Language} & \textbf{LoC} & \textbf{Stars} & \textbf{Domain} \\
\midrule
Chromium & C++, JavaScript & 23M & 23.2K & Browser, Javascript Runtime \\
Firefox & C++, JavaScript & 18.9M & 11.6K & Browser, Javascript Runtime \\
LLVM & C, C++ & 9.9M & 23K & Compiler \\
Hermes & C, C++, JavaScript & 3.9M & 10.8K & Javascript Runtime \\
OpenSSL & C & 3.5M & 29.8K & Cryptography \\
FFmpeg & C & 1.3M & 58.1K & Multimedia \\
Wasmtime & Rust & 520K & 17.8K & WebAssembly Runtime \\
Redis & C & 241K & 73.5K & Distributed Database \\
SQLite & C & 156K & 9.2K & Database \\
FreeType & C & 140K & 784 & Font Library \\
QuickJS & C & 94K & 2.8K & Javascript Runtime \\
Memcached & C & 60K & 14.1K & Distributed Database \\
\bottomrule
\end{tabular}
}
\vspace{-0.8em}
\end{table}

\parabf{Bug Finding Setup.}
To find new bugs in each system and demonstrate \tool's effectiveness,
we first run \simplereporter to generate candidate bug reports.
We ask the reporter to use time windows ranging from 6 months to 2 years for different systems based on their historical bug-fixing activity.
Then we run \tool on each candidate bug report to validate it.
\tool can support arbitrary agent scaffolds, and we select two representative agents: Claude Code\cite{claudecode} and Codex\cite{codex}, and couple them with a mixture of models, including Claude \sonnet{4.5 \& 4.6}, \opus{4.5 \& 4.6}, and \gptcodex{5.3}.
Agents can use the default tools of the scaffolds, such as editor, search, web search, and bash execution.
We only restrict the use of interactive tools, such as \texttt{AskUserQuestion} in Claude Code, to enable autonomous validation.
We construct the Docker image for each system, including the source code and a built binary.

\parabf{Baselines.}
Given the diverse types of large-scale systems \tool can handle, there are no existing dedicated \poc generation tools that can support such heterogeneous settings.
Thus, we choose the state-of-the-art general-purpose \llm coding agents as baselines: {Claude Code~\cite{claudecode} with \opus{4.5}} and {Codex~\cite{codex} with \gptcodex{5.2}}.
For a fair comparison, we evaluate two variants of \tool, each with the same agent scaffold and model as the baseline.
Namely, \toolclaude for Claude Code and \toolcodex for Codex.
All agents are given the same bug reports and task descriptions, and run in the same Docker environments.

\subsection{Generality and Practicality}
\label{sec:eval:bug_finding}

\begin{table}[t]
\centering
\caption{New bugs detected by \tool.}
\label{tab:bugs}
\small\setlength{\tabcolsep}{4pt}
\begin{tabular}{lccccc}
\toprule
\textbf{Project} & \textbf{Total} & \textbf{Confirmed} & \textbf{Fixed} & \textbf{Test} & \textbf{Pending} \\
\midrule
Firefox & 32 & 32 & 23 & 16 & \textendash \\
OpenSSL & 17 & 16 & 14 & 5 & 1 \\
LLVM & 11 & 4 & 4 & 4 & 7 \\
Chromium & 9 & 7 & 5 & 4 & 2 \\
FreeType & 9 & 9 & 9 & \textendash & \textendash \\
SQLite & 9 & 8 & 8 & 2 & 1 \\
Wasmtime & 9 & 9 & 8 & 7 & \textendash \\
FFmpeg & 8 & 8 & 8 & \textendash & \textendash \\
Redis & 6 & 6 & 4 & 4 & \textendash \\
QuickJS & 5 & 5 & 5 & 3 & \textendash \\
Hermes & 3 & 1 & 1 & 1 & 2 \\
Memcached & 3 & 3 & 3 & \textendash & \textendash \\
\midrule
\textbf{Total} & \textbf{121} & \textbf{108} & \textbf{92} & \textbf{46} & \textbf{13} \\
\bottomrule
\end{tabular}
\vspace{-0.8em}
\end{table}

\begin{figure}[t]
\centering
\includegraphics[width=\columnwidth]{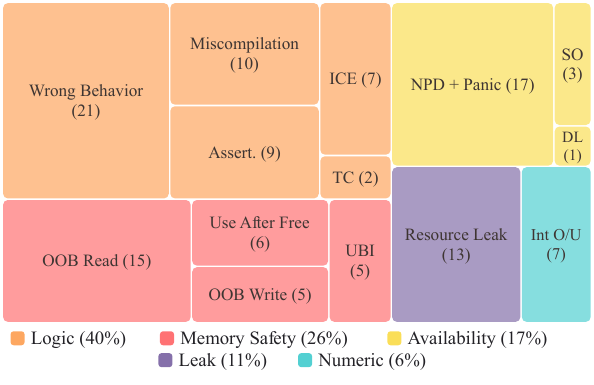}
\caption{Bug type distribution. \textnormal{ICE: internal compiler error, TC: type confusion, NPD: null pointer dereference, SO: stack overflow, DL: deadlock, OOB: out-of-bounds, UBI: use-before-initialization, Int O/U: integer over/underflow, Assert.: assertion failure.}}
\label{fig:bug_types}
\vspace{-0.9em}
\end{figure}

\parabf{Scaling Bug Detection.}
Over the campaign, \simplereporter produced roughly \campaignAllPoCsApproximate candidate reports, the vast majority of which were noise (e.g., non-bugs or duplicates).
\tool enabled just three of the authors to triage this entire pool down to \numTotalBugs genuine new bugs across all \numSUT systems, as shown in \autoref{tab:bugs}.
Of these, \numBugsConfixed are confirmed by developers, \numBugsFixed are fixed, and the rest are still pending; in addition, \numPoCAsTest generated \pocs have been adopted as official regression tests.
For comparison, Anthropic's recent vulnerability-finding campaign \cite{ant2026cvd} hired \emph{six security research companies} to triage \emph{1900} potential findings, leading to 97 vulnerabilities being fixed.
While it targets security vulnerabilities specifically, the contrast in manual effort (six companies for 1900 findings vs.\ three authors for \campaignAllPoCsApproximate) shows that \tool makes large-scale triage manageable for a small team, enabling bug detection at a scale otherwise infeasible in practice.

\parabf{System Generality.}
We observe that \tool is effective across all \suts, implemented in diverse languages such as C, C++, Rust, and JavaScript (\autoref{tab:projects}), without customizing any rules or patterns or relying on system-specific dependencies to navigate and interact with each system.
This flexibility even allows \tool to automatically find bugs in domain-specific languages (DSL).
For example, LLVM uses a DSL called TableGen\cite{llvm_tablegen}.
Within the NVIDIA backend's instruction definitions expressed in this DSL, \tool identified a sign-inversion bug that leads to silent miscompilation of \texttt{sitofp} in any SM90+ (Hopper) kernel that converts a 1-bit integer to \texttt{bfloat}.
Existing works that rely on specific parsers or static analyzers would struggle to find such bugs.

\parabf{Bug Type Generality.}
\autoref{fig:bug_types} shows the type distribution of the bugs detected by \tool.
\tool can automatically handle diverse kinds of bugs, spanning logic errors, robustness issues, and security vulnerabilities.
\tool finds many logic errors that are overlooked by existing tools and developers.
For example, \tool discovers a spec-compliance bug in Firefox's JavaScript engine SpiderMonkey\cite{spidermonkey} for the Await feature.
During the fix, Firefox developers noticed that this behavior is never tested by the official specification test suite.
Without overly focusing on security-only issues, \tool shows great potential in upholding general software quality.
At the same time, \tool is also able to find many memory safety issues and security vulnerabilities.
Two of the bugs found by \tool have been assigned CVE numbers.
For example, \tool detects a use-before-initialization bug in Firefox's support for SVG fonts.
This vulnerability was fixed within one day and assigned a CVE number CVE-2026-2806.

\begin{figure}[t]
\centering
\includegraphics[width=\columnwidth]{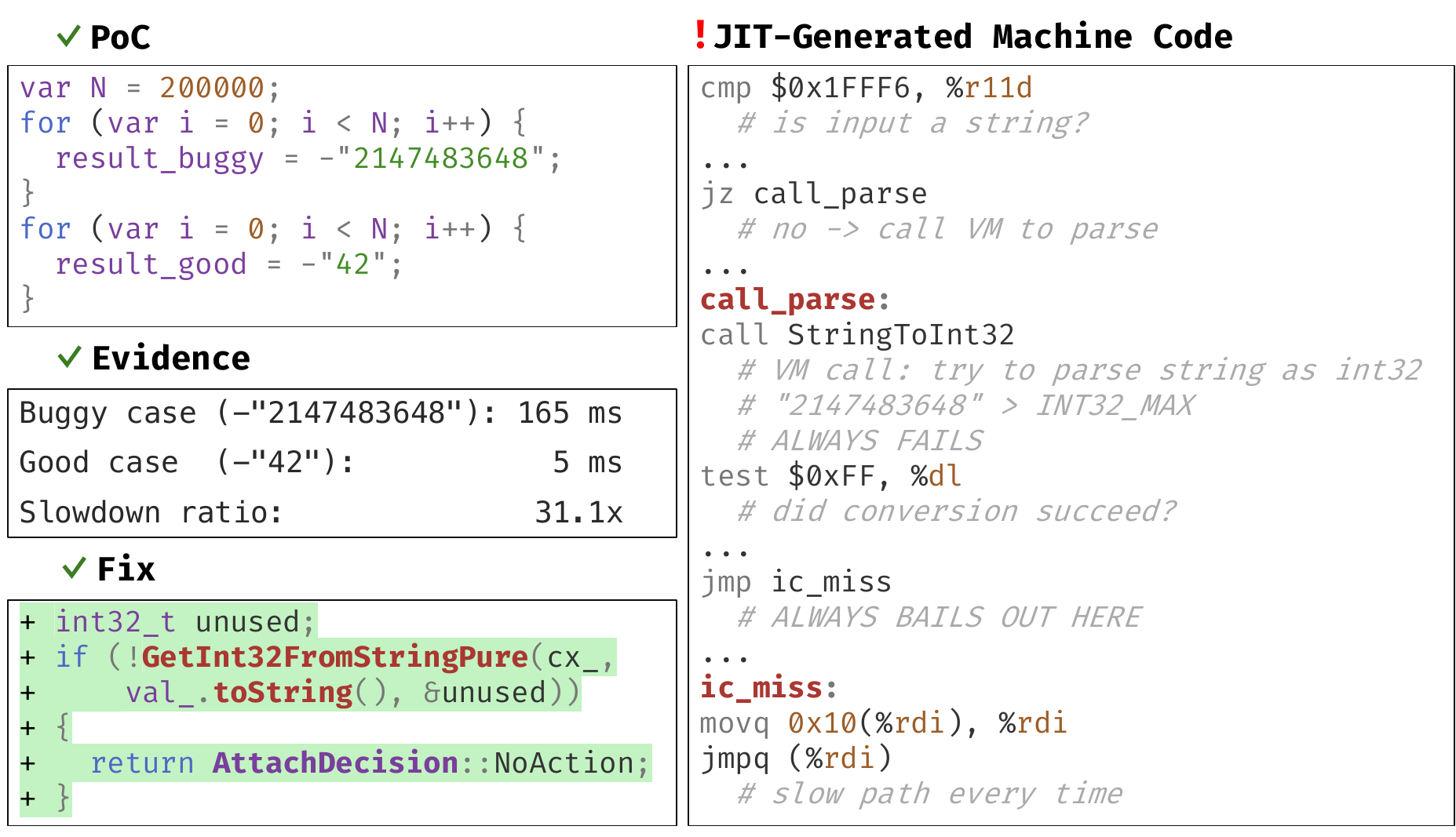}
\caption{Bug example in Firefox's JavaScript JIT engine.}
\label{fig:code_example}
\vspace{-0.8em}
\end{figure}

\begin{table*}[t]
\centering
\caption{Comparison of \tool against baseline agents.}
\label{tab:main_comparison}
\small\setlength{\tabcolsep}{4.5pt}
\setlength{\aboverulesep}{0pt}\setlength{\belowrulesep}{0pt}\renewcommand{\arraystretch}{1.25}
\begin{tabular}{llcc>{\columncolor{gray!12}}c@{\hspace{9pt}}cc>{\columncolor{gray!12}}c@{\hspace{15pt}}>{\columncolor{gray!12}}cc@{\hspace{9pt}}>{\columncolor{gray!12}}cc@{\hspace{15pt}}c}
\toprule
& & \multicolumn{6}{c}{Positive} & \multicolumn{4}{c}{Negative} & \\
\cmidrule(lr){3-8} \cmidrule(lr){9-12}
& & \multicolumn{3}{c}{LLM Reported} & \multicolumn{3}{c}{Human Reported} & \multicolumn{2}{c}{LLM Reported} & \multicolumn{2}{c}{Human Reported} & \\
\cmidrule(lr){3-5} \cmidrule(lr){6-8} \cmidrule(lr){9-10} \cmidrule(lr){11-12}
\textbf{Agent} & \textbf{Model} & $\bot$ & Invalid & Valid & $\bot$ & Invalid & Valid & $\bot$ & Invalid & $\bot$ & Invalid & \textbf{Avg.\,Cost} \\
\midrule
\textit{Gold} & --- & 0 & 0 & 48 & 0 & 0 & 48 & 48 & 0 & 48 & 0 & --- \\
\hdashline
Claude Code &  \opus{4.5} & 0 & 22 & 26 & 2 & 29 & 17 & 1 & 47 & 3 & 45 & \$2.42 \\
\textbf{\toolclaude} &  \opus{4.5} & 9 & 1 & 38 & 19 & 5 & 24 & 42 & 6 & 40 & 8 & \$5.02 \\
\midrule
Codex &  \gptcodex{5.2} & 2 & 23 & 23 & 7 & 17 & 24 & 6 & 42 & 8 & 40 & \$1.26 \\
\textbf{\toolcodex} &  \gptcodex{5.2} & 12 & 1 & 35 & 21 & 1 & 26 & 45 & 3 & 47 & 1 & \$2.61 \\
\bottomrule
\end{tabular}
\vspace{-0.8em}
\end{table*}

\parabf{\poc Format Generality.}
\tool can generate \pocs in various formats, since the ways to interact with \numSUT different systems and the ways bugs manifest vary vastly.
\autoref{fig:code_example} shows a bug detected by \tool in Firefox's JavaScript JIT engine.
The bug occurs when a string operand used with a unary operator is converted to a number by the JIT compiler.
This bug causes the optimized machine code to always fail, forcing execution to fall back to the slower interpreter.
This silent miscompilation does not raise any error or produce any incorrect results.
\tool automatically generates a compact performance test that proves the bug through execution-time measurements, as shown on the left of \autoref{fig:code_example}.
This enables developers to identify and fix the issue quickly, without manually inspecting JIT-generated machine code, as illustrated on the right of \autoref{fig:code_example}.
Without depending on specific format requirements, \tool can automatically identify the most appropriate \poc format for different bugs.

\parabf{Developer Feedback.}
During our bug-finding process, we received extensive positive feedback from developers.
For example, Firefox maintainers reached out to learn about \tool and commented:
\begin{quotebox}
``Thank you for reporting all these correctness bugs in SpiderMonkey! High quality bug reports and we noticed many great findings :)''
\end{quotebox}
Wasmtime maintainers expressed interest in potentially applying \tool in their CI pipeline and noted:
\begin{quotebox}
``Thanks for a really high quality bug report! \ldots\ plus the nice regression test \ldots''
\end{quotebox}
The maintainers of FreeType suggested running \tool on unfuzzed code paths to find additional interesting bugs not covered by fuzzing.
This highlights its potential as a complement to fuzzing thanks to its ability to easily cover more code paths.
For several bugs in OpenSSL, developers explicitly requested that the \poc generated by \tool be added as a regression test.
Notably, two reported bugs were classified as intended behavior, but our reports helped developers revise the documentation; these two are not included in \autoref{tab:bugs}.

\subsection{Validation Effectiveness}
\label{sec:eval:vanilla}

As described in \autoref{sec:background} and \autoref{sec:background:problem-setting}, a key challenge of using \poc test generation is to correctly reject invalid reports while still generating valid \pocs for true bug reports.

\parabf{Setup.}
Existing datasets commonly used for the \poc generation task\cite{wang2025cybergym, lee2025secbench, zhang2025cybench} are insufficient for two reasons.
First, they contain only true bugs and thus cannot fully evaluate \tool's effectiveness.
Second, they are usually restricted to specific bug types or \poc formats, so they cannot evaluate \tool's generality.
To overcome the above challenges, we sample reports generated by our \simplereporter and also manually collect more reports from other real users to construct a comprehensive dataset of both valid and invalid reports.
We construct the dataset with reports generated by \simplereporter during the bug finding campaign with a cutoff time of 2026-02-10.
By that time, \simplereporter generated \campaignTotalRuns reports covering 7 of the selected \suts (Chromium, Firefox, OpenSSL, FFmpeg, SQLite, Wasmtime, Redis).
We include all \datasetLLMPos true bug reports from these reports.
We further randomly sample \datasetLLMNeg reports that are rejected by \tool, and manually verify that they are indeed invalid. Given the reports produced by our \simplereporter may not be representative of bugs found by other detection techniques,
we further collect \datasetHumanPos confirmed bug reports and \datasetHumanNeg rejected reports from other users for the same 7 systems in their bug tracking systems.
In total, this dataset contains \datasetTotal bug reports.

We evaluate the two variants of \tool and baseline agents on this dataset.
Each agent is given the same bug report and run in the same Docker environment with full Bash access.
For each bug report, we ask the agent to generate a valid \poc or report impossible otherwise.
For each generated \poc, we manually verify its validity.
We require the \poc to trigger the bug by interacting with the corresponding system and to show concrete evidence, such as a sanitizer crash or a difference between two execution traces.
We mark \pocs as invalid for the following scenarios:
(1) the \poc is a standalone program that only mimics the \sut behavior without actually exercising the buggy code paths;
(2) the \poc invokes the \sut's internal APIs, relies on invalid input data, or requires unrealistic setup, whereas the bug can be triggered through normal interaction with public interfaces instead;
(3) the \poc encounters execution errors that are not related to the bug.

\parabf{Results.}
\autoref{tab:main_comparison} shows the comparison results.
Reports are grouped by positive (true) and negative (false) bug reports, each further split by source (LLM-reported and human-reported).
For each bug report, an agent can either reject it ($\bot$), produce an invalid \poc, or produce a valid \poc.
The \textit{Gold} row shows the ground-truth outcome for each category.
For the positive reports, \toolclaude is able to generate \evalPosImprovePercentageClaude more valid \pocs than the vanilla \cc,
while \toolcodex generates \evalPosImprovePercentageCodex more valid \pocs than the vanilla \codex.
Notably, both \toolclaude and \toolcodex produce near-zero invalid \pocs, whereas about half of candidate \pocs generated by baseline agents are non-functional or invalid \pocs.
For the negative reports, vanilla \cc and \codex almost always claim to have successfully generated \pocs for the false bug reports.
They fail to reject \evalNegativeNotRejectedRateClaude and \evalNegativeNotRejectedRateCodex of the false bug reports, respectively.
This high false positive rate indicates that vanilla agents tend to game the task by generating only plausible \pocs that do not provide any concrete evidence of the bug.
On the contrary, \toolclaude and \toolcodex can correctly reject \evalNegativeRejectRateOurClaude and \evalNegativeRejectRateOurCodex of the false bug reports, respectively.
By coupling a low invalid-\poc rate on true reports with a high rejection rate on false reports, \tool shows its potential as a scalable and reliable validation oracle for \llm-based bug detection.

According to \autoref{tab:main_comparison}, \tool incurs higher costs than baseline agents.
This is because producing a valid \poc typically requires substantially more codebase exploration and multiple rounds of debugging and refinement.
We also conduct a price-matched experiment with \cc by running it twice.
\autoref{tab:price_match} shows that even with a similar budget, the baseline still produces fewer valid \pocs.
More importantly, it produces \priceMatchBaselineGenerated candidate \pocs that all require human triage, \priceMatchTriageRatio more manual inspection effort than the \priceMatchOursGenerated \pocs \tool surfaces.
This introduces a significant downstream human cost that far outweighs the cheap token savings and defeats the purpose of using \poc generation to scale bug detection in the first place.

\begin{table}[t]
\centering
\caption{Price-matched comparison.}
\label{tab:price_match}
\small\setlength{\tabcolsep}{6pt}
\setlength{\aboverulesep}{0pt}\setlength{\belowrulesep}{0pt}\renewcommand{\arraystretch}{1.25}
\begin{tabular}{l c c c c c}
\toprule
\textbf{Agent} & \textbf{$\bot$} & \textbf{Invalid} & \textbf{Valid} & \textbf{\pocs} & \textbf{Avg.\,Cost} \\
\midrule
Claude Code\textsubscript{1st} & 6 & 143 & 43 & 186 & \$2.42 \\
Claude Code\textsubscript{2nd} & 12 & 141 & 39 & 180 & \$2.25 \\
\cmidrule(lr){1-6}
Claude Code\textsubscript{both}$^{\dagger}$ & 4 & 132 & 56 & 366 & \$4.67 \\
\midrule
\textbf{\toolclaude} & 110 & 20 & 62 & 82 & \$5.02 \\
\bottomrule
\addlinespace[2pt]
\multicolumn{6}{@{}p{\dimexpr\linewidth-2\tabcolsep\relax}@{}}{%
\scriptsize $\dagger$\,this row presents whether the two runs produce \textbf{any} valid PoC for the Valid column, and any invalid PoC (w/o valid PoC) for the Invalid column.%
} \\
\end{tabular}
\vspace{-0.8em}
\end{table}

In summary, our evaluation shows that \tool reliably rejects invalid bug reports while generating high-quality \pocs for true ones, demonstrating its effectiveness as a validation oracle for scalable, end-to-end \llm-based bug detection.

\subsection{Ablation Study}
\label{sec:eval:components}
We study the contribution of each component in \autoref{sec:approach}.

\begin{table}[t]
\centering
\caption{False reports rejected by each agent.}
\label{tab:agent_ablation}
\small\setlength{\tabcolsep}{3.5pt}
\begin{tabular}{lrrrr}
\toprule
\textbf{Agent} & \textbf{Analyzer} & \textbf{Generator} & \textbf{Checker} & \textbf{Total} \\
\midrule
\toolclaude & 42 (51.2\%) & 29 (35.4\%) & 11 (13.4\%) & 82 \\
\toolcodex & 83 (90.2\%) & 7 (7.6\%) & 2 (2.2\%) & 92 \\
\midrule
\textbf{Average} & 62.5 (71.8\%) & 18 (20.7\%) & 6.5 (7.5\%) & 87 \\
\bottomrule
\end{tabular}
\vspace{-0.8em}
\end{table}

\parabf{Agents}
We study the effectiveness of the three subagents by counting how many false reports each rejects, as summarized in \autoref{tab:agent_ablation}.
On average, each variant rejected \evalAgentFalseTotalAvg false reports, with the analyzer rejecting \evalAgentAnalyzerFalseAvg (\evalAgentAnalyzerFalsePctAvg) after statically analyzing the source code, the generator rejecting \evalAgentGeneratorFalseAvg (\evalAgentGeneratorFalsePctAvg) after attempting to generate and execute a \poc, and the checker rejecting \evalAgentCheckerFalseAvg (\evalAgentCheckerFalsePctAvg) after scrutinizing the \pocs and execution traces.
Although the checker rejects the fewest reports, every false report it catches would otherwise have required substantial human effort to triage.
Overall, each subagent rejects a significant number of false reports and contributes to the effectiveness of \tool.
Note that the split of rejected reports varies across the two variants.
\toolcodex rejects more reports at the analyzer stage whereas \toolclaude rejects more at the generator and checker stages.
This leads to a lower overall cost for \toolcodex as reported in \autoref{tab:main_comparison}.
However, this aggressive early rejection also causes \toolcodex to discard more true reports than \toolclaude.
This reveals that different models exhibit different trade-offs.
In practice, one can mix and match different models across subagents based on the actual use case.

\begin{table}[t]
\centering
\caption{\poc generation for 96 positive reports w/ and w/o KB.}
\label{tab:knowledge_comparison}
\small\setlength{\tabcolsep}{6pt}
\setlength{\aboverulesep}{0pt}\setlength{\belowrulesep}{0pt}\renewcommand{\arraystretch}{1.2}
\begin{tabular}{l c >{\columncolor{gray!12}}c}
\toprule
 & \textbf{w/o KB} & \textbf{w/\hphantom{o} KB} \\
\midrule
Valid \pocs & 56 & 65~{\scriptsize($16\%\uparrow$)} \\
Unique & 3 & 12 \\
Avg.\ Cost & \$5.02 & \$4.67~{\scriptsize($7\%\downarrow$)} \\
\midrule
Tool Calls & 55.2 & 44.1~{\scriptsize($20\%\downarrow$)} \\
Bash Cmds & 32.7 & 21.9~{\scriptsize($33\%\downarrow$)} \\
Redundant Re-reads & 3.6 & 1.8~{\scriptsize($49\%\downarrow$)} \\
\bottomrule
\end{tabular}
\vspace{-0.8em}
\end{table}

\parabf{Knowledge Base.}
\label{sec:eval:components:kb}
We also evaluate the contribution of the self-evolving knowledge base (KB).
We first compare \tool's performance with and without it.
Due to limited resources, we rerun the \toolclaude variant with and without KB only for the \datasetTotalTrue true bug reports, since invalid reports normally do not reach the generator and KB usage.
Similarly, since the analyzer also does not use KB and introduces significant noise, we rerun \poc generation with the analyzer disabled.
As shown in \autoref{tab:knowledge_comparison}, \tool generates \kbCompKbValid valid \pocs with KB, a \kbCompValidPct improvement over the \kbCompNokbValid valid \pocs generated without KB.
The KB variant uniquely solves \kbCompKbUnique reports versus only \kbCompNokbUnique for the no-KB variant.
It also lowers the average cost per report (\kbCompKbCost vs.\ \kbCompNokbCost) by \kbCompCostPct.

We further inspect how the KB improves generation by collecting tool-usage statistics over the \kbEffNumBugs \pocs both variants successfully generate, summarized in the lower half of \autoref{tab:knowledge_comparison}.
With the KB, \tool issues \kbEffToolCallsPct fewer tool calls and \kbEffBashPct fewer bash commands to generate the same \pocs.
At the same time, the KB variant makes fewer redundant re-reads of the same file, dropping from \kbEffNokbRereadsAvg to \kbEffKbRereadsAvg per \poc.
This indicates that the generator takes advantage of the accumulated knowledge to avoid re-discovering recurring setup and debugging steps from scratch.

We also collect usage statistics over the whole knowledge base.
Across the evaluation dataset, \tool accumulates \kbTotalItems knowledge items that are used \kbTotalUsages times in total, with each generation drawing on \kbAvgItemsPerGen items on average.
The generator rates each used item, yielding an average rating of \kbAvgRating (median \kbMedianRating), indicating that it indeed finds the accumulated knowledge useful.
We also found the top 5 most-used items relate to recurring tasks, such as setting up sanitizers and debugging tools, and using the \sut's testing framework.
The items further undergo \kbTotalVersionUpdates self-evolving version updates, each raising an item's average rating by \kbAvgRatingImprovementPerUpdate.
This shows \tool can continuously accumulate and reuse knowledge to generate \pocs more effectively and efficiently on the fly.

\section{Threats to Validity}\label{sec:limitation}

\parabf{Internal.}
One threat to validity is that agents, either baseline or \tool, might cheat the evaluation in \autoref{sec:eval:vanilla} by accessing online resources that contain working \pocs or related discussion.
We cannot disable network access because many complex \poc generation tasks indeed require access to online resources, such as documentation or additional dependencies.
To mitigate this, we first use \llms to search for evidence of cheating behaviors in the trajectories.
Furthermore, two of the authors manually checked all trajectories and did not find any cheating.
We acknowledge that this is not a perfect solution.
For example, the agents may still hide the cheating behavior within a complex script.
However, we believe our mitigation is effective enough to reduce the risk of cheating and thus ensure the validity of the evaluation.

\parabf{External.}
The main external threat to our claims of generality lies in the target systems that we apply \tool on.
These systems might not be representative of the entire software ecosystem.
To mitigate this, we specifically include mature industrial software systems that exercise common software engineering practices.
These systems are heavily reviewed, tested, and fuzzed.
They span many different programming languages and domains.
The systems have also been extensively used in software testing and analysis work\cite{hazimeh2020magma, klees2018evaluating, metzman2021fuzzbench, wang2025cybergym, will2012intoverflow}.
We believe our mitigation is sufficient to demonstrate the generality of \tool.

\section{Conclusion}

We presented \tool, a universal \poc generation framework that can serve as a scalable validator for \llm-based bug detection.
We conducted extensive experiments on real-world software systems to demonstrate the effectiveness of \tool.
Across \numSUT large-scale, critical software systems, \tool detected \numTotalBugs new bugs/vulnerabilities, with \numBugsConfixed already confirmed by developers and \numBugsFixed fixed. \numPoCAsTest \pocs have also been adopted as official regression tests.

\balance
\bibliographystyle{IEEEtran}
\bibliography{references}

\end{document}